\documentclass[reprint,amsmath,amssymb,prl,superscriptaddress,nofootinbib]{revtex4-2}

\newcommand{\ra}{\rightarrow}

\newcommand{\End}{{\rm End}}

\newcommand{\CC}{{\mathbb C}}
\newcommand{\ZZ}{{\mathbb Z}}
\newcommand{\RR}{{\mathbb R}}

\newcommand{\HH}{{\mathbb H}}

\newcommand{\cH}{{\mathcal H}}

\newcommand{\Cl}{{C\ell}}

\newcommand{\fotimes}{\mathbin{\widehat\otimes}}

\newcommand{\cN}{\mathcal N}

\newcommand{\be}{\begin{equation}}
\newcommand{\ee}{\end{equation}}

\usepackage{amsmath,amssymb,amsfonts, amsthm}
\usepackage{physics}
\usepackage{dsfont}

\usepackage{graphicx}
\usepackage{dcolumn}
\usepackage{bm}

\usepackage{xcolor}

\usepackage{hyperref}

\begin{document}

\preprint{APS/123-QED}

\title{Supersymmetric Boundaries of One-Dimensional Phases of Fermions beyond SPTs}

\author{Alex Turzillo}
\email{alex.turzillo@mpq.mpg.de}
\affiliation{Max-Planck-Institut f\"ur Quantenoptik, Hans-Kopfermann-Stra{\ss}e 1, 85748 Garching, Germany}
\affiliation{Munich Center for Quantum Science and Technology, Schellingstra{\ss}e 4, 80799 M\"unchen, Germany}

\author{Minyoung You}
\email{minyoung.you@apctp.org}
\affiliation{Asia Pacific Center for Theoretical Physics (APCTP),
Pohang 790-784, South Korea}

\date{December 8, 2020}

\begin{abstract}

It has recently been demonstrated that protected supersymmetry emerges on the boundaries of one-dimensional intrinsically fermionic symmetry protected trivial (SPT) phases. Here we investigate the boundary supersymmetry of one-dimensional fermionic phases beyond SPT phases. Using the connection between Majorana edge modes and real supercharges, we compute, in terms of the bulk phase invariants, the number of protected boundary supercharges.

\end{abstract}

\maketitle

Much recent interest has been paid to topological phases of matter and their classification. Simplest among them are invertible phases, which are essentially trivial in their bulks but host topologically protected phenomena on their boundaries \cite{Kitaevtalk}.\footnote{Their name is due to invertible phases having inverses under the operation of stacking \cite{Kitaevtalk, Kitaevtalk2}. They have also been called short-range entangled (SRE) phases, though there is disagreement around the use of this term.} The notion of topological phases may be enriched by considering systems invariant under a global symmetry and restricting deformations to symmetric ones. A large class of symmetry enriched invertible phases is given by symmetry protected trivial (SPT) phases -- those which would belong to the trivial phase in the absence of the protecting symmetry \cite{ChenGuWenone,ChenGuWentwo}. Beyond SPT phases are invertible phases that remain topologically distinct even without symmetry.

A particularly interesting class of invertible phases is those of fermions in one dimension, which includes, for example, the topological superconductor, whose boundary Majorana modes distinguish it from the trivial superconductor \cite{KitaevMajorana}. The absence of noninvertible topological order in one dimension means that every one-dimensional indecomposable phase without symmetry breaking is invertible. The problem of classifying and characterizing these phases in the symmetry enriched context was solved over a decade ago by Fidkowski and Kitaev in Ref. \cite{FK10}. Roughly speaking, for the group $G_b$ of symmetries modulo fermion parity, phases are described by three topological invariants: group cochains $\alpha\in C^2(G_b;U(1))$ and $\beta\in C^1(G_b;\ZZ_2)$, subject to constraints and equivalences, and a value $\gamma\in\ZZ_2$. The invariant $\gamma$ measures whether a phase supports an even or odd number of modes on each boundary. Since even numbers of modes may be gapped out by interactions that disrespect the symmetry, a value of $\gamma=0$ indicates that the phase is SPT. The invariants $\alpha$ and $\beta$ may also be understood in terms of protected features of the boundary physics.

One such feature is the projectivity of the symmetry action on the boundary. The $2$-cochain $\alpha$ always measures the projectivity of the $G_b$ action on the boundary, while the meaning of the $1$-cochain $\beta$ depends on the value of $\gamma$. For an SPT phase, $\beta(\bar g)$ encodes whether the boundary action of a symmetry $\bar g\in G_b$ commutes or anticommutes with fermion parity. For a phase that is not SPT, $\beta(\bar g)$ instead encodes commutation of the $\bar g$ action with the central fermionic boundary mode $\Gamma$.

While the bulk phase invariants constrain the projective action of the symmetry on the boundary, they do not fix it absolutely. In particular, the number and statistics of the boundary degrees of freedom on which the symmetry acts is not determined by the invariants. For example, a system in the trivial phase may have zero energy degrees of freedom on its boundaries, yet it belongs to the same phase as the trivial system obtained by gapping out these degrees of freedom. This is true of $k=2$ copies of the nontrivial class D Majorana chain or $k=8$ copies of the class BDI Majorana chain, for example. For an example of a system with nontrivial order, consider a stack of $k=4$ class BDI Majorana chains. This system has four Majorana zero modes on each of its boundaries, yet it belongs to the same phase as the system obtained by partially gapping out the boundaries in a way that leaves a Kramer's doublet of bosonic zero modes. Despite the collection of boundary modes of a system not being an invariant of the system's phase, it is still possible to make statements about \emph{protected} modes. If the constraint imposed by the phase invariants on the projective action of the symmetry is such that a minimal collection of modes is necessary to realize the constraint, these modes will be present in every system belonging to the phase.

The perspective of characterizing a phase by its boundary degrees of freedom is suited to studying the supersymmetry that emerges on the boundary. Supersymmetry is the existence of parity-odd operators called supercharges that satisfy the relations of a supersymmetry algebra \cite{WittenMorse}. Supercharges on a zero-dimensional space are closely related to Majorana modes, and the supersymmetry algebra to their Clifford algebra \cite{CliffSUSY}. Supersymmetry is especially interesting when it emerges on the boundaries of topological phases because its supercharges may be protected by bulk topological invariants, meaning the supersymmetry requires no fine-tuning. This occurs, for example, in a recent letter of Prakash and Wang \cite{PW20}, which constructs two real supercharges on the boundary of some one-dimensional SPT phases and argues that they are protected by the SPT invariant $\beta$.

The purpose of the present Letter is to investigate the possibility of protected supersymmetry on the boundaries of one-dimensional fermionic phases beyond SPT phases. Ultimately, we are able to determine the number of protected supercharges as a function of the bulk phase invariants. These findings are stated as \hyperlink{r1}{\textbf{Result 1}} and \hyperlink{r2}{\textbf{Result 2}} in the text. We find examples of phases that protect arbitrarily many boundary supercharges. The case of class BDI superconductors is also discussed in detail. In an appendix, we discuss the distinct phenomenon (also called ``supersymmetry'') wherein a symmetry is represented on a boundary or defect as a parity-odd operator.

\vspace{5mm}

\noindent \textbf{\textit{Symmetries and invariants:}} We begin by reviewing the symmetry groups and topological invariants of one-dimensional phases of fermions. The invariants are understood in terms of the action of the symmetry on an algebra of boundary operators. This algebra will feature in later sections when we ask whether it contains modes generating a supersymmetry algebra.

A fermionic symmetry group $G_f$ has a central involution $p$ called fermion parity. Centrality of $p$ is the condition that there are no parity-odd symmetries in the bulk realization of the symmetry. In general, the extension
\begin{equation}
    \ZZ_2^f\ra G_f\xrightarrow{b}G_b
\end{equation}
of the quotient $G_b=G_f/\ZZ_2^f$ by the subgroup $\ZZ_2^f=\{1,p\}$ does not split as a product group
\begin{equation}\label{productgroup}
    G_f=G_b\times\ZZ_2^f~.
\end{equation}
However, in order for $G_f$ to be realized as the symmetries of a fermionic system that is not SPT, it must split as a product group of this form \cite{FK10,spinTQFT,TurzilloYou}.\footnote{The realization of $G_f$ as a product group requires a physically meaningful choice of subgroup isomorphic to $G_b$.} Whether a symmetry of $G_b$ is represented unitarily or anti-unitarily on the physical state space is encoded by a map
\be x:G_b\ra\ZZ_2^T~.\ee
The triplet $(G_f,p,x)$ specifies the symmetry class.

Fermionic phases of symmetry class $(G_f,p,x)$ have a classification, originally due to Ref. \cite{FK10} (see also Refs. \cite{spinTQFT,TurzilloYou,Bultinck_2017}) in terms of three invariants
\be \alpha\in C^2(G_b;U(1))~,\quad\beta\in C^1(G_b;\ZZ_2)~,\quad\gamma\in\ZZ_2~, \ee
subject to certain constraints and equivalences. In the case of a product group, the invariants $\alpha$ and $\beta$ represent classes in group cohomology twisted by the action where $\bar g\in G_b$ with $x(\bar g)$ inverts the coefficient.

One way of understanding the invariants is in terms of the action of the symmetries on the algebra of operators on the boundary. Here we briefly review how this works, leaving detailed discussion to Refs. \cite{spinTQFT,TurzilloYou}. The invariant $\gamma$ measures whether this algebra is of the form
\begin{align}
&A=\End(U_f)&&\quad(\gamma=0\text{, even, SPT})\qquad\text{or}\nonumber\\
&A=\End(U_b)\otimes C\ell(1)&&\quad(\gamma=1\text{, odd, not SPT})~,\nonumber
\end{align}
where $U_f,U_b$ are vector spaces, $\End$ denotes the algebra of matrices on a space, and $C\ell(1)$ is the complex Clifford algebra with one generator. An algebra (and its corresponding system) is referred to as ``even'' when $\gamma=0$ or ``odd'' when $\gamma=1$. For example, the algebra generated by $N$ Majorana modes is the Clifford algebra $C\ell(N)$, which is isomorphic to an even or odd algebra depending on whether $N$ is even or odd. Systems belong to the same phase in the absence of symmetry if their algebras are related by $\End(U)$ factors \cite{spinTQFT}; this means that $\gamma$ encodes whether a system is SPT. The symmetries act on the algebra as follows, according to the invariants:
\begin{align}
    \gamma=0:\quad&g\cdot M = Q_f(g)MQ_f(g)^{-1}\\
    \gamma=1:\quad&\bar g\cdot M\otimes\Gamma^m\nonumber\\
    &\qquad=(-1)^{\beta(\bar g)m}Q_b(\bar g)MQ_b(\bar g)^{-1}\otimes\Gamma^m\label{oddaction}\\
    &p\cdot M\otimes\Gamma^m = (-1)^mM\otimes\Gamma^m~,
\end{align}
where $Q_f,Q_b$ are projective representations of $G_f,G_b$, and $g\in G_f,\bar g=b(g)\in G_b$. The cocycle measuring the projectivity of $Q_b$ is simply $\alpha$, while that of $Q_f$ is a certain function of $\alpha$ and $\beta$ \cite{spinTQFT,TurzilloYou}, such that $\beta$ may be extracted from $Q_f$ as
the phase in the commutator
\begin{equation}\label{evenbeta}
    PQ_f(g)P=(-1)^{i\pi\beta(\bar g)}Q_f(g)
\end{equation}
of the actions of $g$ and fermion parity $P=Q_f(p)$.

\vspace{5mm}

\noindent\textbf{\textit{Supercharges from zero modes:}} We now establish a connection between Majorana zero modes and real supercharges. This connection has been noted previously elsewhere, for example in Ref. \cite{Behrends_2020}.

Consider a system with Majorana zero modes $\gamma_i$, not necessarily protected. Any set of $N$ modes forms the Clifford algebra $\Cl(N)$. Diagonalize the Hamiltonian of the system as $H=\sum_\mu E_\mu\mathds{1}_\mu$, where $\mathds{1}_\mu$ projects onto the eigenspace labeled by $\mu$. Since the modes have zero energy, they must commute with $H$ and so also with the $\mathds{1}_\mu$. We may define $N$ real supercharges as
\be Q_i=\sum_\mu\sqrt{E_\mu}\mathds{1}_\mu\gamma_i~.\ee
They satisfy the supersymmetry algebra
\be \{Q_i,Q_j\}=\sum_\mu E_\mu\mathds{1}_\mu\{\gamma_i,\gamma_j\}=2\delta_{ij}H~.\ee
Conversely, given an algebra of real supercharges $Q_i$, a Clifford algebra of Majorana zero modes is recovered as
\be \gamma_i=\sum_\mu\frac{1}{\sqrt{E_\mu}}\mathds{1}_\mu Q_i~.\ee

Now suppose the $N$ modes are protected by virtue of living on the boundary of a nontrivial phase. This implies that the supersymmetry is protected as well.

It is worth mentioning a subtlety to the counting of supercharges that arises when $N$ is odd. Given a fermion parity $P$ local to the system, the presence of a single real supercharge $Q_1$ implies there is a second supercharge $Q_2=iPQ_1$ \cite{supercharge}. A similar fact is true of any odd number: fermion parity constructs one additional supercharge to make the total count even. For this reason, systems intrinsic to zero dimensions, where there is a local fermion parity, always have an even number of real supercharges. This is why works like Refs. \cite{alvarez-gaume, ingmar} work with complex -- rather than real -- supercharges as the basic unit. On the other hand, the boundary of an odd one-dimensional phase has no local fermionic parity, so it really has an odd number of supercharges. A nonlocal fermion parity may be constructed as $P=i\Gamma\gamma_\infty$ by adding an extra mode $\gamma_\infty$ to the odd $C\ell(N)$ \cite{FK10, Behrends_2020}. Physically, this ``Majorana mode at infinity'' can be the Majorana mode at the opposite boundary of the Majorana chain \cite{KitaevMajorana, FK10} or a Majorana mode at a far-separated vortex in a two-dimensional topological superconductor \cite{ReadGreen}, for example. One may then speak of the odd number of ``local supercharges'' and the one additional ``nonlocal supercharge'' built from the nonlocal $P$ \cite{Behrends_2020}. In this letter, we will count only the supercharges of the \emph{local} supersymmetry algebra.

\vspace{5mm}

\noindent\textbf{\textit{Counting protected supercharges:}} We now investigate when a phase protects $\cN$ Majorana zero modes on its boundary. As we just saw, this amounts to studying protected boundary supersymmetry of $\cN$ real supercharges. The terms Majorana zero mode and real supercharge are henceforth used interchangeably.

Recall from a previous section that a fermionic system is characterized by an algebra $A$ with a compatible action of the symmetries $G_f$. This data may be interpreted as the algebra of zero energy degrees of freedom on the boundary and how they transform under symmetry. The form of the algebra is constrained by the phase invariants: $\gamma$ determines whether the algebra is of the form $\End(U)$ or $\End(U)\otimes\Cl(1)$, while $\alpha$ (and also $\beta$ in the even case) detects the projectivity of the group action on $U$.

Given an algebra $A$, we ask for the largest $N$, denoted $N(A)$, such that $A$ has a tensor factor decomposition
\be A=B\fotimes\Cl(N)~,\label{factordecomp}\ee
where the hat over $\otimes$ reminds us to use the tensor product graded by fermionic parity, if neither factor is purely even. The number $N(A)$ represents the number of Majorana modes, or real supercharges, present on the boundary of the system. Since we are interested in fermionic rather than bosonic modes, we require that the $C\ell(N)$ factor is not purely even. If $C\ell(N)$ has at least one odd generator $\gamma_j$, any even generator may be replaced by an odd one by the graded isomorphism $\gamma_i'\sim\gamma_i\gamma_j$.

Two systems belong to the same phase if they have the same topological invariants, which is to say that their algebras satisfy the same constraints. We are interested in features protected by the phase; that is, in characteristics of the class $[\alpha,\beta,\gamma]$ of algebras compatible with given values the invariants. The absence or presence of the $\Cl(1)$ factor is a characteristic of the class because it is associated with $\gamma$. The invariants $\alpha$ and $\beta$ provide more flexibility: there may be multiple distinct irreducible projective representations $U$ with the same projectivity class, and so their algebras are associated with systems in the same phase. This means the particular $\End(U)$ factor is \emph{not} a characteristic. In physical terms, a typical zero energy boundary degree of freedom of a system is not a protected feature of the phase, as only some of these are present in every system in the phase. Here we ask, given a phase, what the \emph{protected} zero modes on its boundary are. This amounts to looking at all of the algebras $A$ in the class and asking for the smallest value of $N(A)$:
\be \mathcal{N}(\alpha,\beta,\gamma)=\min_{A\in[\alpha,\beta,\gamma]}N(A)~. \label{protmodes}\ee
In the following, we will compute the numbers $\mathcal{N}(\alpha,\beta,\gamma)$ of protected boundary Majorana zero modes. We begin with odd phases before turning to SPT phases.

\vspace{5mm}

\noindent\textbf{\textit{Odd case:}} Consider the algebra associated with a system in an odd phase. It has the form
\be A=M(d)\otimes\Cl(1)~,\ee
where $M(d)$ denotes the algebra of complex $d\times d$ matrices. The $C\ell(1)$ factor is generated by an odd zero mode $\Gamma$. Our question is whether there exist additional modes.

We claim that the number of odd zero modes is
\be N=2k+1~,\ee
where $k$ is the largest whole number such that $2^k$ divides the degree $d$ of the representation $U$.

To see this, let $l=d/2^k$ and note that
\begin{align}\begin{split}
    M(d)\otimes\Cl(1)&\simeq M(l2^k)\otimes\Cl(1)\\
    &\simeq M(l)\otimes M(2^k)\otimes\Cl(1)\\
    &\simeq M(l)\otimes\Cl(2k+1)~.
\end{split}\end{align}

So far we have counted the modes on the boundary of the system associated with the algebra $A$. But we are interested only in the modes that are protected by the phase to which this systems belongs. Since the value $d$ is the degree of an irreducible representation with projectivity class $\alpha$, the number of protected modes, given abstractly by Eq. \eqref{protmodes}, is the following:

\vspace{3mm}

\noindent\hypertarget{r1}{\textbf{Result 1:}} \emph{The number of real supercharges $\mathcal{N}(\alpha,\beta,1)$ protected by an odd phase with invariant $\alpha$ is exactly $2k+1$, where $k$ is the largest number for which $2^k$ divides $d_\alpha$, the greatest common divisor of the degrees of the irreducible representations with projectivity class $\alpha$.}

\vspace{3mm}

We cannot present a general, explicit formula for $\cN$ because no such expression for $d_\alpha$ is known; however, the mathematics literature contains some limited results that hold at least when $G_b$ contains only unitary symmetries. Upper and lower bounds on $\cN$ follow from the facts that $d_\alpha$ divides the order of $G_b$ and is divided by the order of $\alpha$ in cohomology (cf. Ref. \cite{Berkovich}, corollary VI.3.10 and lemma VI.4.1). More can be said when $G_b$ is a finite Abelian group. In this case, every irreducible representation with projectivity class $\alpha$ has the same degree
\begin{equation}
    d_\alpha = \sqrt{|G|/|K_\alpha|}~,
\end{equation}
where $|\cdot|$ denotes the order and $K_\alpha$ is the subgroup
\begin{equation}
    K_\alpha = \{g\in G\text{ : }\alpha(g,h)=\alpha(h,g)\text{ , }\forall h\in G\}
\end{equation}
(cf. Ref. \cite{Berkovich}, theorem VI.6.6, and Ref. \cite{backhouse}).

For an example that is common in studies of symmetry-enriched phases, consider the group $G_b=\ZZ_n\times\ZZ_n$. Its irreducible projective representations are described by clock and shift matrices on spaces of dimension the order of $\alpha$ in $H^2(G_b;U(1))=\ZZ_n$ \cite{Sylvester}.\footnote{We emphasize that the ``generalized Clifford algebra'' of the clock and shift matrices has to do with the symmetry action protecting the modes and is unrelated to the ordinary Clifford algebra formed by the modes themselves.} Let $n=2^k$ and take $\alpha$ to generate the cohomology group. Then $\mathcal{N}(\alpha,\beta,1)$ is $2k+1$. This class of examples can be used to obtain an arbitrarily large number of protected supercharges.

Another important case is odd phases with trivial $\alpha$, for which the number $\mathcal{N}(0,\beta,1)$ is exactly $1$. To see this, note that if $\alpha$ is trivial, it represents the projectivity class of the trivial representation, which has degree $d=1$ and so $k=0$, corresponding to one real supercharge. Since this mode is protected (as in any odd phase) and is the only mode in one system (the one with trivial representation), it is the only protected mode for the phase.

Let us also illustrate how our formula applies to phases of symmetry class BDI -- that is, $G_f=\ZZ_2^T\times\ZZ_2^f$. The construction of supercharges for each $\nu = 0,\ldots,7$, has been carried out in Ref. \cite{Behrends_2020}, and we recover their counting. Using the dictionary, given in Refs. \cite{FK10,TurzilloYou}, between the number of layers $\nu$ and the invariants $\alpha,\beta,\gamma$, our formalism recovers this count of supercharges:
\begin{itemize}
    \item In the phases with $\gamma=\nu\mod 2=1$, there is at least one protected supercharge.
    \item When $\nu=1,7$, $\alpha$ is trivial and so there are no more protected supercharges. We have a total of $\cN = 1$.
    \item When $\nu=3,5$, $\alpha$ is nontrivial and the smallest irreducible projective representation (the Kramer's doublet) has degree $d=2$, for a total of $\mathcal{N}=3.$
\end{itemize}
We also could have recovered this counting from looking at the ``minimal'' algebras given by the real superdivision algebras: see the table in Ref. \cite{TurzilloYou}, and note that $\Cl_{1,0}\RR$ and $\Cl_{0,1}\RR$ each have a single generator while $\mathbb{H}\otimes\Cl_{0,1}\RR$ and $\mathbb{H}\otimes\Cl_{1,0}\RR$ each have three.

\vspace{5mm}

\noindent\textbf{\textit{Even case:}} Consider the algebra associated with a system in an SPT phase. It has the form
\be A=M(a|b)~, \ee
where $M(a|b)$ denotes the graded algebra of matrices on the graded vector space $\CC^{a|b}$.

We begin by arguing that, since the projective action of $G_f$ on $U=\CC^{a|b}$ is irreducible (by assumption), the grading is either purely even $b=0$ (when $\beta$ is trivial) or equal $a=b$ (when $\beta$ is nontrivial). Recall the interpretation \eqref{evenbeta} of the invariant $\beta$ in an SPT phase. It measures whether a symmetry acts on $U=\CC^{a|b}$ as an odd (invertible) operator. Note that the grading is equal precisely when such an operator exists. Therefore, a nontrivial $\beta$ implies that the grading is equal. On the other hand, consider the case where $\beta$ is trivial, meaning the symmetry acts as even operators, i.e. within the even-even and odd-odd blocks. By irreducibility, one of these (up to isomorphism, the odd-odd block) must vanish; therefore, the grading is purely even. This proves our claim.

Next observe that
\be M(a|a)\simeq M(a)\otimes\Cl(2)~.\label{cl2factor}\ee
This means that, if $M(a|b)$ has equal grading $a=b$, it has at least two odd modes.\footnote{We emphasize that $\Cl(k)$ denotes the Clifford algebra without the purely even grading. The algebra $M(2|0)$ is $\Cl(2)$ only as an ungraded algebra, and its generators are even modes, which are unrelated to supercharges.} Also note that, for $k>0$,
\begin{align}\begin{split}
    M(c|d)\fotimes\Cl(2k)&\simeq M(c|d)\fotimes M(2^{k-1}|2^{k-1})\\
    &\simeq M((c+d)2^{k-1}|(c+d)2^{k-1})~.
\end{split}\end{align}
This implies the converse: that, if $M(a|b)$ has unequal grading $a\ne b$, it contains no odd modes.

From this we may conclude the following, which was first proved by Prakash and Wang:

\vspace{3mm}

\noindent\textbf{Lemma} (cf. Ref. \cite{PW20})\textbf{:} \emph{If $\beta$ of an SPT phase is trivial, the number of protected real supercharges $\mathcal{N}(\alpha,0,0)$ is $0$, while, if $\beta$ is nontrivial, $\mathcal{N}(\alpha,\beta,0)$ is at least $2$.}

\vspace{3mm}

Next, we state an SPT analog of Result 1. As discussed briefly in a previous section and more extensively in Refs. \cite{spinTQFT,TurzilloYou}, the projectivity class of the group action on $U$ is a certain lift $\omega$ of $\alpha$ to from $G_b$ to $G_f$ that combines the data of $\alpha$ and $\beta$.\footnote{On a technical note, readers familiar with Refs. \cite{spinTQFT, TurzilloYou} will recall that $\alpha$ is subject to an equivalence arising from the freedom $\mu$ to change the splitting map. This, however, does not affect our Result 2 because $\omega$ is invariant under this transformation. Similarly, in the odd case, since $\cN$ is independent of $\beta$, it is not affected by $\mu$.} In terms of this class $\omega$, we have

\vspace{3mm}

\noindent\hypertarget{r2}{\textbf{Result 2:}} \emph{The number of real supercharges $\cN(\alpha,\beta,0)$ protected by an SPT phase with invariants $\alpha, \beta$ is $0$ if $\beta$ is trivial; if $\beta$ is nontrivial, it is exactly $2k$, where $k$ is the largest number for which $2^k$ divides $d_\omega$, the greatest common divisor of the degrees of the irreducible representations with projectivity class $\omega$.}

\vspace{3mm}

The proof uses the reasoning from the odd case. If $\beta$ is nontrivial, the algebra has the form \eqref{cl2factor}. Then
\be M(a|a)\simeq M(l)\otimes\Cl(2m+2)~, \ee
where $a=l2^m$. The degree $d$ of the representation is $2a$, so $d=l2^k$ for $k=m-1$. This means that the number of supercharges is $2k$, where $k$ is the largest number for which $2^k$ divides the degree of the representation. To complete the argument, recall Eq. \eqref{protmodes}, which says that the number of protected supercharges is the minimum of the number of supercharges over all systems in the phase.

As an example, consider $G_f=\ZZ_{2^k}\times\ZZ_{2^k}$. By the argument presented in the odd case, we see that the number of protected supercharges $\mathcal{N}(\alpha,\beta,0)$ is $2k$ for any phase with $\alpha, \beta$ such that $\omega$ generates the cohomology group. As before, this class of examples can be used to obtain an arbitrarily large number of protected supercharges.

For SPT phases with trivial $\alpha$, the number $\cN(0,\beta,0)$ is $0$ if $\beta$ is trivial and exactly $2$ if $\beta$ is nontrivial. To see this, note that if $\alpha$ is trivial, the only projectivity in the action of $G_f$ comes from commutators of $Q_f(g)$ with $P$. Therefore, $P$ may be represented as $\sigma_z$ and each $Q_f(g)$ with $\beta(\bar g)=0,1$ as $\mathds{1}, \sigma_x$, respectively. This representation has degree $d=2$, so the number of protected supercharges is at most $2$. Applying the lemma completes the proof.

Let us look again at the case of class BDI, this time for the even phases. We recover the counting of Ref. \cite{Behrends_2020}:
\begin{itemize}
    \item When $\nu=0,4$, $\beta$ is trivial and so there are no protected supercharges.
    \item When $\nu=2$, $\beta$ is nontrivial while $\alpha$ is trivial, so there are $\cN=2$ protected supercharges.
    \item When $\nu=6$, both $\beta$ and $\alpha$ are nontrivial. These are compatible with a representation of degree $d=2$ (given by the real graded algebra $\Cl_{0,2}\RR$), so there are still only $\cN=2$ protected supercharges.
\end{itemize}
Again, we could have looked at real superdivision algebras, where $\RR$ and $\HH$ are purely even while $\Cl_{2,0}\RR$ and $\Cl_{0,2}\RR$ each have two odd generators.

\vspace{5mm}

\noindent\textbf{\textit{Summary and Outlook:}} We have found that supersymmetry emerges without fine-tuning on the boundaries of a broad class of one-dimensional symmetry-enriched phases of fermions -- both SPT and beyond. The lack of fine-tuning is by virtue of its protection by the topological invariants $(\alpha,\beta,\gamma)$ that characterize the bulk phase. For each phase, we computed the number $\cN(\alpha,\beta,\gamma)$ of protected real supercharges. Our results extend the recent discovery that intrinsically fermionic SPT phases support at least $\cN=2$ protected boundary supersymmetry \cite{PW20}.

This work opens a line of investigation into the consequences of the emergent supersymmetry for the Sachdev-Ye-Kitaev (SYK) models that arise on the boundaries of one-dimensional fermionic phases that have been many-body localized. The quantum chaotic eigenspectra of these models have been shown to encode information about the bulk topological invariants \cite{PhysRevB.95.115150}. In the setting of bulk phases of symmetry class BDI, which support the eightfold way of SYK models on their boundaries, this feature of the spectra and related properties of dynamical correlation functions were shown to be constrained by supersymmetry \cite{Behrends_2020}. Our results raise the possibility of understanding the connection between supersymmetry and these phenomena in a much broader class of phases. It would also be interesting to study how our work generalizes to phases of quantum matter in higher dimensions. We leave these questions and others for another day.

\vspace{5mm}

\noindent\textbf{\textit{Acknowledgements:}} A.T. acknowledges support from the Max Planck Institute of Quantum Optics (MPQ) and the Max Planck Harvard Research Center for Quantum Optics (MPHQ). M.Y. is supported by an appointment to the JRG Program at the APCTP through the Science and Technology Promotion Fund and Lottery Fund of the Korean Government and by the Korean Local Governments - Gyeongsangbuk-do Province and Pohang City.

\bibliography{bibi}

\appendix

\section{Appendix: Parity-odd symmetries}

A distinct phenomenon that also goes by the name ``supersymmetry'' is the existence of a parity-odd symmetry. This can emerge in a topological phase when a symmetry, despite commuting with fermion parity in the bulk, is projectively represented as an odd operator on boundaries or defects. For $P$ the (possibly non-local) fermion parity operator, the condition reads
\begin{equation}\label{fermsym}
    V(g)P=(-1)PV(g)~.
\end{equation}
This phenomenon was previously explored in Ref. \cite{TRsuperconductor}, where authors Qi, Hughes, Raghu, and Zhang study a two-dimensional class DIII superconductor and find that time-reversal symmetry anticommutes with fermion parity on vortices; they refer to this phenomenon as supersymmetry. Supersymmetry of this sort has been known since Fidkowski and Kitaev to occur in one dimension, where the boundaries of certain phases are acted on by parity-odd symmetries \cite{FK10}. As we have already noted, this is precisely what $\beta$ measures in SPT phases.

%When the term ``supersymmetry'' is used in this appendix, it refers to this notion, which we note is only incidentally related to the supersymmetry of the main text. 

This appendix is dedicated to understanding the extent to which parity-odd symmetries, protected and not, are present in one-dimensional invertible phases that are not SPT. The odd number of modes complicates the structure of the boundary state space and changes the interpretation of $\beta$. We find that, in contrast with SPT states, for which the boundary condition is essentially unique (and either hosts parity-odd symmetries or not), invertible states that are not SPT may host multiple distinct boundary conditions parametrized by $1$-cocycles $\delta\in Z^1(G_b;\ZZ_2)$. Just like $\beta$ in the SPT case, the new datum $\delta$ encodes the commutation of symmetries with fermion parity; the crucial difference is that $\delta$ is a choice rather than a bulk phase invariant. When $\delta$ is nontrivial, there exists a parity-odd symmetry. We also investigate the action, given by stacking, of SPT phases on invertible phases and their boundary conditions.

\vspace{5mm}

\noindent\textbf{\textit{Boundary conditions for invertible phases:}} A boundary condition is characterized by the action $T$ of the algebra, with a compatible projective action $V(g)$ of the symmetries, on a space of states $\cH$.

For an SPT phase, whose algebra is of the form $A=\End(U_f)$, the boundary condition is essentially unique. The only indecomposable action of the algebra with
\be T(M)T(N)=T(MN)~,\ee
for $M,N$ matrices in $\End(U_f)$, is as matrices on the space $\cH=U_f$. To see that the projective symmetry action is also unique, apply the compatibility condition
\begin{equation}\label{equimod}
    T(g\cdot M)=V(g)T(M)V(g)^{-1}~,
\end{equation}
which implies
\begin{align}\begin{split}
    V(g)T(M)V(g)^{-1}&=T(g\cdot M)\\
    &=T(Q_f(g)MQ_f(g)^{-1})\\
    &=T(Q_f(g))T(M)T(Q_f(g))^{-1}~.
\end{split}\end{align}
Therefore, up to an unphysical phase, the symmetry acts as $V(g)=T(Q_f(g))=Q_f(g)$.

On the other hand, the algebra of boundary operators of an odd phase is of the form $A=\End(U_b)\otimes C\ell(1)$. This algebra also has a unique (graded) indecomposable action -- this time on the state space
\be \cH=U_b\otimes\CC^2~, \ee
with the $\CC^2$ factor acted on by $C\ell(1)$. One solution for the projective symmetry action is given by
\begin{equation}
    V(p)=P~,\quad V(\bar g)=P^{\beta(\bar g)}T(Q_b(\bar g)\otimes 1)\label{nonsusybc}
\end{equation}
where $P$ is an involution on $\cH$ that acts trivially on $U_b$ and anticommutes with the action of $\Gamma$ on $\CC^2$. Note that $P$ is not part of the algebra action but rather generates $\End(\cH)$ together with the algebra action. But this is not the only solution to the conditions \eqref{oddaction} and \eqref{equimod}: for each $1$-cocycle $\delta\in Z^1(G_b;\ZZ_2)$, a solution is given by
\begin{equation}\label{boundarycondition}
    V(p)=P~,\quad V(\bar g)=P^{\beta(\bar g)}T(Q_b(\bar g)\otimes\Gamma^{\delta(\bar g)})~.
\end{equation}
For example, in the simplest case of $A=C\ell(1)$ acting on $\cH=\CC^2$, there is a binary choice for each $\bar g\in G_b$ as to whether it acts as $P^{\beta(\bar g)}$ or $P^{\beta(\bar g)}T(\Gamma)$. To see that this is the complete solution, note that since $P$ and $T(A)$ generate $\End(\cH)$, any map on $\cH$ has the form
\begin{equation}
    V(g)=P^{\epsilon(g)}T(N(g)\otimes\Gamma^{\delta(g)})~.
\end{equation}
To condense notation, let $t(g)\in\{0,1\}$ denote the value $(\bar g,0)\mapsto 0$, $(\bar g,p)\mapsto 1$. Then impose compatibility:
\begin{align}
    (-1)^{(\beta(\bar g)+t(g))m}&T(Q_b(\bar g)MQ_b(\bar g)^{-1}\otimes\Gamma^m)\nonumber\\\begin{split}
    &=T(g\cdot M\otimes\Gamma^m)\\
    &=V(g)T(M\otimes\Gamma^m)V(g)^{-1}\end{split}\\
    &=(-1)^{\epsilon(g)m}T(N(g)MN(g)^{-1}\otimes\Gamma^m)~.\nonumber
\end{align}
Assuming the algebra action $T$ is faithful, it follows that $\epsilon(g)=\beta(\bar g)+t(g)$ and $N(g)$ is related to $Q_b(\bar g)$ by an unphysical phase; meanwhile, $\delta$ is so far unconstrained. To see that $\delta$ is closed, require that $V(g)$ forms a projective representation. Finally, to see that $\delta$ reduces from $G_f$ to $G_b$, impose $V(p)=P$, which means $\delta(p)=0$. We have recovered eq. \eqref{boundarycondition} as a general boundary condition.

The commutator of boundary symmetry actions
\be PV(g)P^{-1}=(-1)^{\delta(\bar g)}V(g)~ \label{paritydelta}\ee
means that parity-odd symmetries may occur and is measured by $\delta$; in fact, the solution \eqref{nonsusybc} with trivial $\delta$ is the unique boundary condition without parity-odd symmetries. In contrast with the parity-odd symmetries on the boundaries of SPT phases, those described here are not an invariant of the phase, as they depends on the choice of independent boundary datum $\delta$. In other words, the parities of $V(g)$ are not enforced by the anomaly of the boundary theory. Nevertheless, since the datum $\delta$ is discrete, it is still the case that no fine-tuning is required, under the physical assumption that smooth deformations of the microscopic boundary Hamiltonian do not cause the $\delta$ of the low energy state space $\cH$ to jump. One question is whether $\delta$ is detectable by a physical observable, and the answer depends on which observables one allows. Using fermion parity, which is not realized in terms of local operators, the commutator \eqref{paritydelta} detects $\delta$.

\vspace{5mm}

\noindent\textbf{\textit{Stacking of boundary conditions:}} Two systems may be stacked together to form a composite system. Since the stacking operation is compatible with phase equivalence, it gives an abelian group structure to the set of fermionic invertible phases. This group has been described in terms of the invariants $\alpha$, $\beta$, and $\gamma$ \cite{spinTQFT,TurzilloYou,Bultinck_2017}.

On the level of the invariant $\gamma$, the stacking law reads
\be\gamma_{12}=\gamma_1+\gamma_2~.\ee
This means that SPT phases form an index two subgroup of the invertible phases. A stack of SPT phases remains an SPT phase, and stacking gives SPT phases a free action on invertible phases, more generally.

In the case of SPT phases, the stacking law says that the invariant $\beta$ behaves under stacking as
\begin{equation}\label{betastack}
    \beta_{12}=\beta_1+\beta_2~.
\end{equation}
Since parity-odd symmetries on the boundaries of SPT phases are controlled by $\beta$, this rule may be interpreted as computing whether a symmetry $g$ is odd on the composite boundary from whether it is on the parts. It may also be observed directly from the commutator of $p$ and $g$ for the composite symmetry action
\begin{equation}\label{stackaction}
    V_{12}(g)=V_1(g)\fotimes V_2(g)~.
\end{equation}
An immediate consequence of this fact is that parity-odd symmetry on the boundaries of two SPT phases with the same $\beta$ is destroyed by stacking the phases together.

The law for stacking an SPT phase with an odd phase reads the same as \eqref{betastack}; however, in this context, $\beta_2$ no longer determines the odd boundary condition, nor does it have the interpretation as measuring parity-odd symmetry. Determining the data of the composite boundary requires looking at the composite action \eqref{stackaction}, which has\footnote{As an alternative to computing the commutator of the $p$ and $g$ actions, one may rewrite $V_{12}(g)$ in the standard form \eqref{boundarycondition}, by use of a certain (graded) isomorphism \cite{spinTQFT}. Then $\delta_{12}$ may be read off from the exponent of $\Gamma$.}
\be \delta_{12}=\beta_1+\delta_2~.\ee
Combined with the law for stacking the phase invariants (realized on the boundaries as anomalies), this rule defines a free action of boundaries of SPT phases on boundaries of phases that are not SPT. According to this action, the anomaly $\beta$ is shifted by the same cocycle $\beta_1$ as the parameter $\delta$.

The stacking of boundaries of two odd phases requires more care than the other cases. Na\"ively computing the commutator of the $p$ and $g$ boundary symmetry actions defined by eq. \eqref{stackaction} would suggest that $\beta_{12}$ is given by $\delta_1+\delta_2$. However, this cannot be correct: the $\delta$'s are independent choices of boundary conditions, while $\beta$ is an invariant of the bulk phase subject to the stacking law, which in the odd-odd case reads
\begin{equation}\label{betastackx}
    \beta_{12}=\beta_1+\beta_2+x~.
\end{equation}
This false paradox is dissolved by carefully accounting for the assumptions underlying the definitions of the invariants. The stacked algebra is the (graded) tensor product
\begin{align}\begin{split}
    A_{12}=A_1\fotimes A_2&=\End(U_{b,1}\otimes U_{b,2})\otimes C\ell(2)\\
    &\simeq\End(U_{b,1}\otimes U_{b,2}\otimes\CC^2)~.
\end{split}\end{align}
It acts on the stacked boundary state space
\be \cH_{12}=\cH_1\otimes\cH_2=U_{b,1}\otimes U_{b,2}\otimes\CC^4 \ee
as two copies of its indecomposable action. These copies are exchanged by the fermion parity operators $P_1$ and $P_2$. Since $A_{12}$ is an even algebra, the proper definition of $V_{12}(g)$ is not $V_1(g)\fotimes V_2(g)$, which is not contained in $T(A_{12})$ as it exchanges the copies; rather, $V_{12}(g)$ is the element of $T(A_{12})$ whose conjugation action agrees with that of $V_1(g)\fotimes V_2(g)$ on the subalgebra $T(A_{12})\subset\End(\cH_{12})$ fixing the copies. The solution \cite{spinTQFT,TurzilloYou}
\begin{align}\begin{split}
    V_{12}(g)=(-i)^{t(g)}T(Q_{b,1}&(\bar g)\otimes Q_{b,2}(\bar g)\\
    &\otimes\Gamma_1^{\beta_2(\bar g)+t(g)}\Gamma_2^{\beta_1(\bar g)+t(g)})~,
\end{split}\end{align}
is unique up to an unphysical phase. With the symmetry action on the composite boundary in hand, the commutator of $g$ and $p$ may be computed to recover the $\beta_{12}$ of eq. \eqref{betastackx} on each copy. The parity-odd symmetry on the full $\cH_{12}$ vanishes for the simple reason that the minus signs of the two copies cancel each other. Notably, the odd boundary data $\delta_1$ and $\delta_2$ do not contribute to the even composite boundary. This stacking rule means that parity-odd symmetry can appear on (each half of) the boundary obtained by stacking two boundaries that lack parity-odd symmetry.

Because of the properties of odd-odd stacking, the rule for stacking boundaries of arbitrary invertible phases is nonassociative and there is no natural unit for the data $\delta$. This is unsurprising since one does not expect the anomalous boundary systems to behave well under an operation that changes the anomaly. (From the perspective of unwinding phases by symmetry extension, changing the anomaly amounts to changing the symmetry class of the boundary \cite{Wang_2018, Prakash_2018, prakash2020unwinding}.) Nevertheless, the action of boundaries of SPT phases on boundaries of arbitrary invertible phases, despite changing the anomaly, is well-behaved.

\end{document}